\documentclass[10pt,letterpaper]{article}
\usepackage{opex3}

\def\aap{ A\&A}
\def\apjs{ Astrophys.\ J.\ }
\def\pasp{ PASP}

\begin{document}

\title{Fast computation of Lyot-style coronagraph propagation}

\author{R. Soummer$^{*1,2}$, L. Pueyo$^3$,\\ A. Sivaramakrishnan $^{1,2,4}$, and R. J. Vanderbei$^{5}$}

\address{$^{1}$Department of Astrophysics, American Museum of Natural History, New York, NY 10024\\
$^2$ Center for Adaptive Optics, University of California, Santa Cruz, CA 95064\\
$^3$Mechanical and Aerospace Engineering, Princeton University, Princeton, NJ 08544\\
$^4$Department of Physics and Astronomy, Stony Brook University, Stony Brook, NY 11794\\
$^5$Operations Research and Financial Engineering, Princeton University, Princeton, NJ 08544\\
*corresponding author: \email{ rsoummer@amnh.org}}



\begin{abstract}
We present a new method for numerical propagation through Lyot-style coronagraphs using finite occulting masks. Standard methods for coronagraphic simulations involve Fast Fourier Transforms (FFT) of very large arrays, and computing power is an issue for the design and tolerancing of coronagraphs on segmented Extremely Large Telescopes (ELT) in order to handle both the speed and memory requirements. 
Our method combines a semi-analytical approach with non-FFT based Fourier transform algorithms. It enables both fast and memory-efficient computations without introducing any additional approximations. Typical speed improvements based on computation costs are of about ten to fifty for propagations from pupil to Lyot plane, with thirty to sixty times less memory needed. Our method makes it possible to perform numerical coronagraphic studies even in the case of ELTs using a contemporary commercial laptop computer, or any standard commercial workstation computer.\\
\end{abstract}




\section{Introduction} \label{sect:intro}
The field of high contrast imaging has expanded rapidly in the past few years, driven by the prospect of science enabled by the direct detection and characterization of extrasolar planets and faint circumstellar disks. 
Several observatories have launched studies and development of such projects, e.g. Gemini Planet Imager (GPI) \cite{MGP06}, SPHERE \cite{BMM06}, or HiCIAO \cite{THT06}.
In the future, Extremely Large Telescopes (ELT) \cite{MTD06,Clo07,VKB07} will provide the higher angular resolution necessary to study more distant planet forming regions, and also faint old giant planets. The study of Earth-like planets will probably have to wait for space-based instruments \cite{GAB06,TT07}.
Ground-based imaging requires the association of coronagraphy and extreme adaptive optics (ExAO), which can be studied both theoretically and numerically \cite{SKM01,AS04,CBB06,GPK06,SFA07}.

In this paper we study the numerical modeling of Lyot type coronagraphs \cite{L32}, including Apodized Pupil Lyot Coronagraphs (APLC) \cite{ASF02,SAF03,S05,SPF07}, or phase masks \cite{RR97,SDA03}. These coronagraphs consist of the succession of binary filters (pupil, occulting mask, Lyot stop).
The method we describe does not provide any significant improvement in the case of infinite-size focal plane masks \cite{RRB00,AVB01,KT02,FPS05}, which we do not discuss further here. 
Under the classical approximations of Fourier Optics \cite{Goo96}, a Fourier Transform (FT) relationship exists between the field amplitude at two successive planes of the coronagraph.
The pupil, with its finite support, is not band-limited and so cannot be perfectly sampled according to the Shanon-Nyquist sampling theorem \cite{Bra99}. The same problem pertains to the occulting mask in the focal plane.
A remedy is to impose very fine sampling both in pupil and focal planes, in order to represent small features (e.g. segmentation or secondary mirror support structures, occulting spot).
Note that Nyquist sampling of the focal plane with two pixels per resolution element $(\lambda/D)$, is not sufficient here. Indeed, the small occulting mask has a typical size of about $5\lambda/D$, and would not be well represented by a disk ten pixels in diameter.
This is even more problematic for phase masks of size $\sim\lambda/D$ \cite{RR97,SDA03,FSA07}. 
Significant oversampling of the PSF is therefore required for sufficient fidelity in coronagraphic modeling.

This two-fold sampling requirement in two Fourier-conjugate planes is not only a severe practical problem, but raises a fundamental problem associated with the uncertainty principle. 
In order to preserve complete information in the FT, a fine sampling in one domain means that the corresponding FT must be calculated to a very high frequency in the reciprocal domain. This is achieved optically when forming successive pupil and focal plane images.
Numerically, this can be reproduced using Fast Fourier Transforms (FFT), but this can lead to overwhelming demands on memory and computing power. 
In Sec.\ref{Sec:classicalMethod} we show that this fundamental limitation makes classical FFTs poorly suited for Lyot-type coronagraphic computations.
In Sec.\ref{sect:semianal} we analyze the analytical formalism of coronagraphic propagation, and conclude that a partial FT algorithm providing arbitrary sampling in a limited area is more appropriate for coronagraphy. Several methods can be used for this purpose \cite{BS01,RSR69}, and we detail a matrix-based FT \cite{Smi07}, which can be implemented easily in any high-level language.

Comparisons between classical propagation with FFTs and our semi-analytical method show a considerable gain both in speed and memory requirements.
This therefore opens new possibilities for coronagraphic studies hitherto impossible without large computer cluster facilities. Among them, the study of extremely fine sampling of the pupil plane in the case of ELTs, of extremely fine sampling of the occulting plane mask, or of end-to-end simulations of an ExAO coronagraph to simulate a long exposure, are achievable with standard desktop equipment.

\section{Lyot-type Coronagraphs} \label{sec:Corono}

\begin{figure}[htbp]
\center
\resizebox{0.75\hsize}{!}{\includegraphics{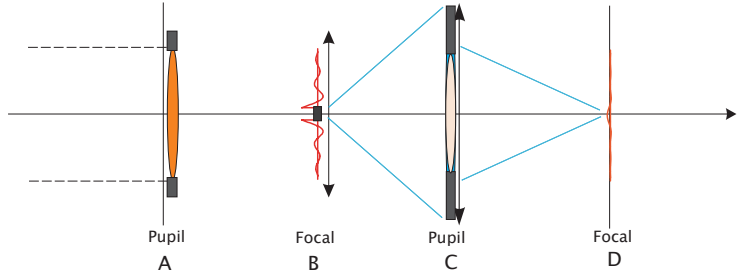}}
 \caption{Illustration of the four coronagraphic planes: the pupil corresponds to Plane A (possibly apodized). A focal masks (hard-edged, or phase mask) is placed in the focal plane B, and a Lyot stop (possibly undersized) in plane C.}
 \label{fig:layout}
 \end{figure}

In this section we recall the general formalism of Lyot-type coronagraphs, following the notation of Aime \textit{et al.} \cite{ASF02} and Soummer \textit{et al.} \cite{SAF03,SPF07}.
The general layout is given in Fig.\ref{fig:layout}. The setup consists of an ensemble of apodizers, masks and stops in four successive planes $A$,$B$,$C$,$D$, respectively, where $A$ is the entrance aperture, $B$ is the focal plane with the occulting mask, $C$ is an image of the entrance aperture where a pupil mask called Lyot stop is placed and $D$ is the final image plane.
We will consider the usual approximations of paraxial optics \cite{Goo96}, and that the optical layout is properly designed to cancel the quadratic phase terms associated with Fresnel propagation, so that a FT relationship exists between two successive planes. 
The telescope aperture function with the position vector $\textbf{r}=(x,y)$ is denoted by $P(\textbf{r})$ (index function equal to 1 inside the aperture $\mathcal{P}$). This aperture can be apodized by a function $\Phi(\textbf{r})$. Note that these functions do not have to be radial. 
A mask of transmission $1-\epsilon M(\textbf{r})$ is placed in the focal plane. $M$ is the index function that describes the mask shape $\mathcal{M}$, equal to 1 inside the coronagraphic mask and 0 outside. $L$ is the index function of the Lyot stop. We recall that both the Lyot coronagraph (opaque mask) and the Roddier coronagraph ($\pi$ phase mask) can be described by this common formalism, with $\epsilon=1$ for Lyot and $\epsilon=2$ for Roddier using that $e^{i \pi}=-1$, as detailed in \cite{ASF02,FSA07}.
The field amplitude in the four successive planes are:
\begin{eqnarray}
\Psi_A(\textbf{r})&=&P(\textbf{r}) \Phi(\textbf{r}) \\
\Psi_B(\textbf{r})&=& \widehat{\Psi}_A(\textbf{r})(1-\epsilon \, M(\textbf{r})) \label{planBcart}\\
\Psi_C(\textbf{r})&=& \big(\Psi_A(\textbf{r})-\epsilon\, \Psi_A(\textbf{r})\ast \widehat{M}(\textbf{r})\big)\,L(\textbf{r}) \label{planCcart} \\
\Psi_D(\textbf{r})&=& \big(\widehat{ \Psi}_A(\textbf{r}) - \epsilon \,\widehat{\Psi}_A(\textbf{r}) M(\textbf{r})\big) \ast \widehat{L}(\textbf{r}) \label{planDcart}
\end{eqnarray}
In these equations, $ \widehat{ } $ \ denotes the Fourier Transform, $*$ the convolution product, and we assume that the focal lengths of the successive optical systems are identical (if not, an appropriate change of variables leads to the same result). Also, we re-orientate the axis in the opposite direction and omit the coordinate reversal for better legibility here.
Also, we assumed monochromatic propagation in these equations. The effect of a finite spectral bandpass can be added easily to this formalism: in the focal plane, the size of the point spread function is proportional to the wavelength, but the mask (hard-edged or phase) has a fixed size. It can be readily shown \cite{ASF02,SAF03,Aime05b} that changing the wavelength is formally equivalent to changing the mask size. 

\section{Classical numerical propagation and FFTs}\label{Sec:classicalMethod}
In order to compute the coronographic propagation, we are interested in the numerical evaluation of the Fourier integral between each plane. Without loss of generality we consider the case of a one-dimensional signal $f(x)$ with $x \in [-\gamma D/2,\gamma D/2]$ and $f(x)=0$ for $|x| > D/2$, where $\gamma$ is a padding coefficient that will be later related to the resolution of the FT in the image plane. We illustrate the relationship between plane A and plane B.
We first recall classical results of Fourier analysis \cite{BS01}; the sampled Continuous Fourier Transform (CFT) can be written as a Riemann sum:
\begin{eqnarray}
\widehat{F}(u_k) &=& \int_{-\gamma D/2}^{\gamma D/2} f(x) e^{-i 2 \pi x u_k} dx\\
&\simeq& \delta x \sum_{n = 0}^{N_A-1} f(x_n) e^{-i 2 \pi x_n u_k }
\label{Eq:Riemann}
\end{eqnarray}
where $x_n = (n - \gamma N_A/2) \delta x$, $n \in [0, \gamma N_A-1]$, corresponding to the sampling points of $f(x)$, and $u_k = (k - N_B/2) \delta u$, $k \in [0,N_B-1]$ are the sampled values of the Fourier transform. $N_A$ is the number of pixels along the pupil diameter $D$, and $N_B$ is the number of pixels in the focal plane. Note that under the Riemann sum approximation, independent sampling grids can be chosen in both domains.
However, when one chooses the same size $N=\gamma N_A=N_B$ for both arrays, and the integration steps as:
\begin{equation}
\delta x \; \delta u = \frac{1}{N}, \label{Eq:Shannon}
\end{equation}
%
then Eq.~\ref{Eq:Riemann} greatly simplifies to:
\begin{equation}
\widehat{F}(u_k) = \frac{\gamma D}{N} (-1)^{N/2-k}\sum_{n=0}^{N-1} (-1)^n f(x_n) e^{-i 2 \pi k n/N},\label{Eq:FFTsum}
\end{equation} 
where the sum is now a Discrete Fourier Transform (DFT), which can be computed very efficiently using FFT algorithms.
The number of floating point operations (flops) in a radix-2 Cooley-Tukey FFT of size $N$ is $5 N \,\log_2 N$. It is generally assumed that this is also an approximation for the number of operations in any complex FFT \cite{FFTW05}.

\begin{figure}[htbp]
\center
\resizebox{\hsize}{!}{\includegraphics{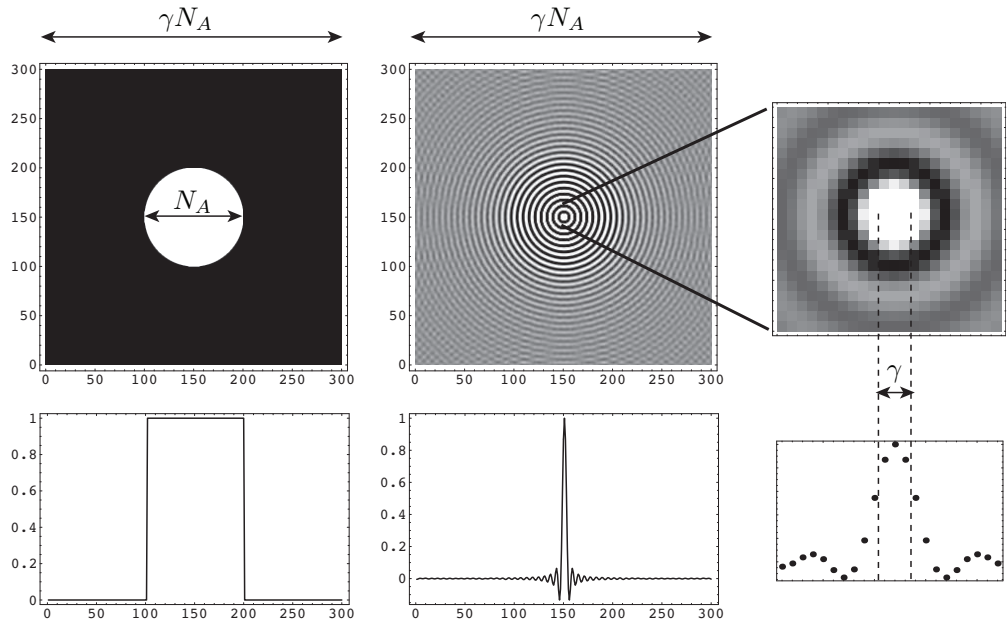}}
 \caption{Sampling relationship between pupil (left column) and image plane (center), using FFTs.
 The padded pupil plane array is $\gamma$ times larger than the actual pupil (here $\gamma = 3$), and its Fourier transform features three pixels per unit of angular resolution, as shown in the zoomed image of the core (right). In this case a $4\sim5$ resolution element mask would not be sufficiently sampled. A general consensus for coronagraphic calculations is to use $\gamma =6$ or $\gamma=8$. }
 \label{fig:PSFsampling}
 \end{figure}

It is critical to realize that this possibility of using FFTs is obtained at the expense of a fixed relationship between the focal plane sampling and the padding factor, forced by Eq.~\ref{Eq:Shannon}: 
$\delta u = \lambda /(\gamma D)$, as illustrated on Fig.~\ref{fig:PSFsampling}. 
Under this condition, the occulting mask of $m$ resolution elements is therefore sampled with $\gamma m$ pixels. Because stellar coronagraphs use very small masks (typically $4\sim5 \lambda/D$ for an APLC, and $\lambda/D$ for a PM or DZPM), decent sampling of these masks (at the very least a few tens of pixels) imposes large zero padding factors $\gamma$. A general consensus for coronagraphic calculation is to use $\gamma=6$ or $\gamma=8$.
The memory and computational speed problems associated with classical coronographic algorithms stem from the requirement of finely sampling the image plane mask. This can lead to prohibitively large padding factors for some applications which can benefit for particularly fine sampling, such as tip-tilt tolerancing \cite{LS05}, study of atmospheric differential refraction effects, mask ellipticity or roughness studies.

A common method to mitigate these mask sampling problems is to use a gray-pixel approximation, where gray pixels are used at the edge of the mask: their value is defined as the ratio of the area covered by the mask to the total pixel area. This technique corresponds to a slight numerical apodization of the focal plane mask, akin to the use of an anti-alizing filter. 
The same approach can be used for small features in the pupil plane, such as secondary mirror support structures. 
However, there are cases where the gray approximation cannot be used. 
For example, the tolerancing of the effects of the small pupil features (segmentation, spiders \cite{SL05}), and their mitigation by the Lyot Stop cannot be done with gray pixel approximation. Prolate apodizers for APLCs need to be defined without gray pixels in the pupil \cite{SPF07}. In the case of phase masks (Roddier \cite{RR97} or Dual Zone \cite{SDA03}), the gray approximation is meaningless for the phase mask, and large padding coefficients have to be used since these masks are very small (typically one resolution element).
 %

\section{Semi-analytical coronagraphic propagations}\label{sect:semianal}
\subsection{Principle}

As shown in Fig.~\ref{fig:PSFsampling}, FFT methods address the image plane sampling requirement by zero padding the pupil, and mimicking the optical propagation, preserving the complete information between planes.
A much better approach can be derived from the analytical expression of the field in the Lyot plane, simply re-writing Eq. \ref{planCcart} as:
\begin{equation}
\Psi_C(\textbf{r})= \left(\Psi_A(\textbf{r})-\epsilon \; \mathcal{F}\big[\mathcal{F}[\Psi_A(\textbf{r})] M(\textbf{r})\big] \right)\, L(\textbf{r}), 
\label{Eq9}
\end{equation}
where we can readily identify that the first FT of the pupil field amplitude $\mathcal{F}[\Psi_A(\textbf{r})]$ is truncated by the occulting spot $M(\textbf{r})$, and that the second FT is truncated by the Lyot Stop $L(\textbf{r})$.
This means that we are only interested in the knowledge of the FTs \textit{inside} limited areas, viz., the limited occulting mask area, and the limited Lyot stop area.
We can thus completely circumvent the sampling problem by restricting the information of the FTs to these two zones: the semi-analytical approach consists of computing these limited-area FTs numerically, and subtracting the result from the pupil complex amplitude, according Eq. \ref{Eq9}. 

In the case of one-dimensional problems (rectangular apertures \cite{ASF02}, or perfect circular apertures \cite{SAF03,SDA03}), the semi-analytical approach can be applied by calculating directly these two limited-area FTs (or Hankel Transforms for circular apertures) using a one-dimensional numerical integration algorithm. This approach presents some advantages for phase masks calculations \cite{SDA03}, but such direct integrations cannot be used efficiently in the two-dimensional case for computation reasons. 

In the general two-dimensional case, the two limited-area FTs can be calculated using partial FT methods. Several methods exist to calculate such partial FTs, such as the Fractional Fourier Transform or chirp z-transform, \cite{BS01,RSR69}, and we describe a very simple matrix-based method in the next section \cite{Smi07}.
These partial FT methods are intrinsically slower that a FFT if exactly the same computation is performed.
However, since we are only interested in a small number of points within limited areas (occulting mask \textit{and} Lyot stop), we actually replace fast calculations with very large arrays by slow calculations with very small arrays. Note that for $\gamma=8$, only $40\times40$ values of the FT need to be calculated in the first focal plane for a $5\lambda/D$ mask, independently of the size of the pupil n pixels.

\begin{figure}[htbp]
\center
\resizebox{0.9\hsize}{!}{\includegraphics{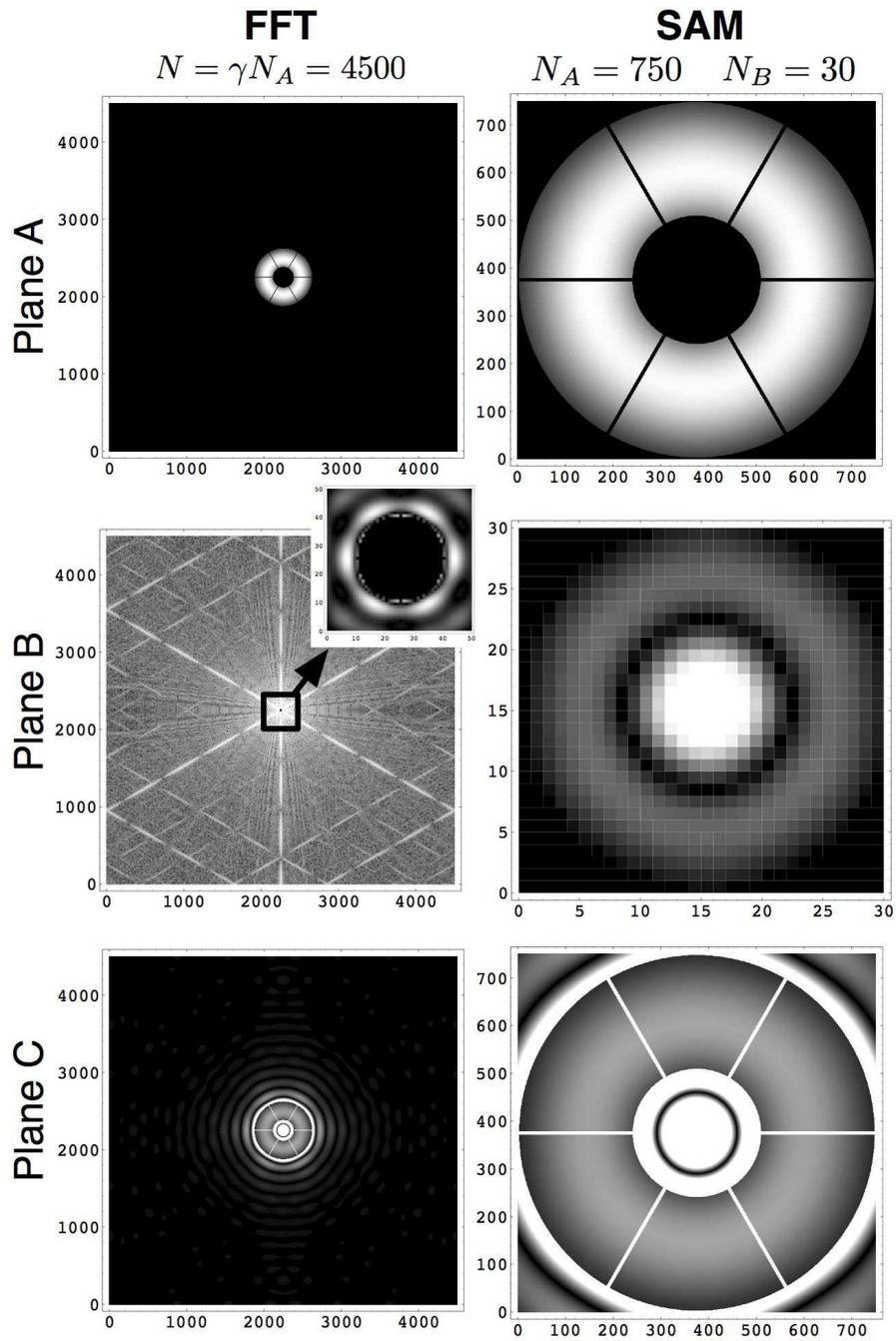}}
 \caption{Computing steps in each plane A,B,C using FFTs with a six-fold zero padding ($4500\times4500$ FFTs), and the semi-analytical method (SAM), for a possible ELT geometry and an APLC.
With SAM, only the points of the FT that are \textit{inside} of the occulting mask are computed as opposed to the FFT method, where the points \textit{outside} the mask are needed. In this example we obtain a speed improvement of about 15 using 36 times less memory.}
\label{fig:IlustrationMethod}
 \end{figure}

The semi-analytical propagation from the pupil to the Lyot plane is performed as follows:
\begin{itemize}
\item Define the pupil field amplitude \textit{without} zero padding as a $N_A\times N_A$ array, where $N_A$ is the diameter of the pupil in pixels.
\item Calculate the FT to an arbitrary fine sampling inside the area \textit{limited} to the occulting mask as a $N_B\times N_B$ array, with e.g. $N_B=40$ for a $5\lambda/D$ mask with $\gamma=8$.
\item Calculate the FT of this previous field amplitude inside an area \textit{limited} to the pupil ($N_A\times N_A$ array).
\item Reverse the spatial axes because two successive FTs restore the function while changing the sign of the variable, and subtract the result from the pupil field amplitude (An inverse FFT can also be used between plane B and plane C to avoid reversing the axis).
\end{itemize}
Note that for convenience all masks (pupil and occulter) are defined with their centers located on the pixel $N_A/2+1$ (or $N_B/2+1$), and that the focal plane mask can be defined using the gray pixel approximation, as for the classical FFT method. 
These computation steps are illustrated in Fig.\ref{fig:IlustrationMethod} and compared to the classical FFT approach. The figure shows the actual arrays that are calculated for both method: note that the semi-analytical method does not use zero-padding in the pupil plane and calculates the focal plane amplitude only \textit{inside} the occulting mask, instead of \textit{outside} in the case of the classical FFT method.

In plane C, the Lyot stop can be equal to the pupil size in the case of an APLC, or more generally undersized to optimize the coronagraphic efficiency according to various possible criteria \cite{SKM01,SPF07}. If the Lyot stop size has been previously chosen, the calculated area can be limited to the Lyot stop itself to accelerate further the computation, but in practice we usually compute the Lyot plane amplitude over the size of the entire pupil.
If necessary, it is straightforward to oversize the calculated region, at the expense of computing efficiency, for example to see the light diffracted outside the pupil.

Avoiding the typical six-to-eight fold zero-padding enables the use of much smaller arrays, than FFT calculations require.
For example, in the case of GPI ($D=8 m$), the secondary mirror support structures are about $1 cm$ wide. Understanding the effects of spiders on coronagraphic performance is important since they appear bright in the Lyot plane and musk be masked out by the Lyot stop \cite{SL05}. 
For example, with 4 pixels per spider, $N_A=3200$ pixels are needed in the pupil. Standard $\gamma=8$ padding would require $25600\times25600$ FFTs. The semi analytical method enables this calculation in a few seconds on a standard desktop computer (Table 2).

It is important to note that the calculations of the partial FTs inside the mask do not correspond to an additional approximation, as it may seem that we discard the information \textit{outside} the focal plane mask. In fact, we do not lose any information because the method is based on the analytical formulation of Eq. \ref{Eq9} which is equivalent to Eq. \ref{planCcart}.

\subsection{Matrix direct Fourier transform}
In this section we describe a simple matrix Fourier transform (MFT), which can be used for the semi-analytical method.
In order to restrict the computation of the FT of the pupil to the image plane size $m$ expressed in resolution elements units $(\lambda/D)$, we choose the sampling step in plane B such that: $du = m/N_B$. We compute the Riemann sum directly using a matrix formulation of Eq.\ref{Eq:Riemann}:
\begin{equation}
\left(\begin{array}{c}\widehat{F}(u_0) \\... \\\widehat{F}(u_k) \\... \\\widehat{F}(u_{N_B-1})\end{array}\right) =
 \left(\begin{array}{ccccc}
 e^{-2 i \pi x_0 u_0} &... & e^{-2 i \pi x_k u_0} &... & e^{- 2 i \pi x_{N_A-1} u_0} \\... &... &... &... &... \\ 
 e^{-2 i \pi x_0 u_k} &... & e^{-2 i \pi x_k u_k} &... & e^{-2 i \pi x_{N_A-1} u_k} \\... &... &... &... &... \\
 e^{-2 i \pi x_0 u_{N_B-1}} &... & e^{-2 i \pi x_k u_{N_B-1}} &... & e^{-2 i \pi x_{N_{A}-1} u_{N_B-1}}
 \end{array}\right) \left(\begin{array}{c} f(x_0) \\... \\ f(x_k) \\... \\f(x_{N_A -1})\end{array}\right),
\end{equation}
which can be rewritten as:
\begin{equation}
\widehat{F}(\mathbf{U}) = e^{-2 i \pi \mathbf{U} \mathbf{X}^{T} } \cdot f(\mathbf{X}),
\end{equation}
where $\mathbf{U} = (u_0.....u_{N_B-1})^{T}$, $\mathbf{X} = (x_0.....x_{N_A-1})^{T}$, and $\exp(\cdot)$ is the element-wise exponential of a matrix.
Two-dimensional FTs can be implemented straightforwardly as follows:
\begin{itemize}
\item Define the four vectors $\mathbf{U} = (u_0.....u_{N_B-1})^{T}$, $\mathbf{X} = (x_0.....x_{N_A-1})^{T}$, $\mathbf{V} = (v_0.....v_{N_B-1})^{T}$, $\mathbf{Y} = (y_0.....y_{N_A-1})^{T}$.
\item The vector elements are $x_k=y_k=(k-N_A/2)\times1/N_A$ and $u_l=v_l=(l-N_A/2)\times m/N_A$, for $k=[0,\dots,N_A-1]$ and $l=[0,\dots,N_B-1]$.
\item The two-dimension FT is obtained by computing the two matrix products:
\begin{equation}
\widehat{F}(\mathbf{U,H}) = \frac{m}{N_A\,N_B}\;e^{-2 i \pi \mathbf{U} \mathbf{X}^{T} }\cdot f(\mathbf{X},\mathbf{Y})\cdot e^{-2 i \pi \mathbf{Y} \mathbf{V}^{T} },
\end{equation}
where the normalization coefficient $m/(N_A\,N_B)$ imposes the conservation of energy according to the Parseval theorem \cite{Bra99}: the energy in the limited-area FT is a fraction of the total energy of the FT, corresponding to the limited area, which was calculated.
\item With this definition, the Fourier transform is centered in a similar fashion to the FFT, with the zero frequency at the pixel $N_B/2+1$.
\end{itemize}
Note that when the conditions of Eq.\ref{Eq:Shannon} are verified, both the FFT and the matrix FT give identical results. This can be used as a numerical check when implementing the technique. This is not surprising since the matrix method is a direct implementation of the Riemann sum, which is also what is computed by a FFT (very efficiently), taking advantage of simplifications for particular sampling conditions (Eq.\ref{Eq:FFTsum}).
With the MFT, the arbitrary sampling $du$ in the focal plane is obtained at the expense of partial information on the FT. Note that in the case of the MFT, $\gamma$ still corresponds to the sampling of the FT (number of pixels per resolution element), but not to the zero-padding coefficient. The semi-analytical formalism frees the computation from the rigid sampling constraints of FFT methods. This provides increased flexibility in the computation, since the number of pixels in the pupil and focal planes can be chosen independently.

In terms of computation cost, the matrix FT involves 2 complex-matrix products of the form:
\begin{equation}
F=E_1\cdot f\cdot E_2,
\end{equation}
where the respective sizes of the three matrix $E_1, f, E_2$ are: $N_B\times N_A$, $N_A\times N_A$, and $N_B\times N_A$. 
We consider the case of a complex FT for generality. Each element of the result in a matrix product is the result of $N_A$ multiplications and $N_A-1$ additions. 
While complex addition requires 2 flops (floating point operations), complex multiplications require 6 flops. Since the two matrices $E_1$ and $E_2$ can be calculated beforehand, in a similar fashion that FFT plans are generated, the total number of operations for the product $f.E_2$ is therefore: $N_A N_B (8N_A-2) \simeq 8 N_A^2 N_B$, assuming $N_A$ large enough.
The number of flops involved in the first complex MFT (pupil to focal) is therefore:
\begin{equation}
n(MFT)=8 (N_A^2 N_B+ N_A N_B^2)-2N_A N_B-2 N_B^2.\label{MFTcost}
\end{equation}
In the particular case where both pupil and image plane have the same array size $N$, it is interesting to note that the number of operations for the matrix method is proportional to $N^3$ and not $N^4$ as for a direct calculation of the DFT. This is because the two successive matrix products take advantage of some redundancy in intermediate calculations. The FFT is as expected more efficient in this case, and the relative number of operations is:
\begin{equation}
\frac{n(FFT)}{n(MFT)}=\frac{5 \,\log_2(N)}{8 N}.
\end{equation}
For example a $2000\times 2000$ FT requires 36 times more operations with a MFT than with a FFT. This calculation, which takes about 0.5s with a FFT on a on a 2GHz Apple G5 (1 Gflop/s), would take 18s with a MFT.
The performance comparison between the FFT propagation method and the semi-analytical method is given in Sec. \ref{Sec:CompCosts}.

\subsection{Fast fractional Fourier transform}
The computation of the FT in a limited area can also be made using a Fractional Fourier Transform. This method is presented by Bailey and Swarztrauber \cite{BS01}, we refer the reader to this work for a complete description of this algorithm. Consider the Riemann sum in Eq.~\ref{Eq:Riemann}, and assume that there is no zero padding $x_n = n \delta x - (N_A \delta x)/2$, $n \in [0,N_A-1]$ with $\delta x = D/ (N_A)$ and $u_k = k \delta u - (N_B \delta u)/2$, $k \in [0,N_B-1]$ with $\delta u = 1 /(\gamma D)$ and $N_B = m \gamma$, where $m$ is the size of the image plane mask in resolution element units $(\lambda/D)$.
Note that here again $\gamma$ does not correspond to an oversize of the pupil but directly to the number of pixels per unit of angular resolution in the image plane. The Riemann sum can be re-written as:
\begin{equation}
\widehat{F}(u_k) = \frac{D}{N} e^{i (\pi/\gamma) (k -N_B/2) }\sum_{n=0}^{N_A} e^{i \pi n N_B /(\gamma N_A)} f(x_n) e^{i 2 \pi k n/(\gamma N_A)}
\end{equation}
which corresponds to a Partial Fourier Transform of the series $g_n = \exp((i \pi n N_B)/(\gamma N_A)) f(x_n) $. This sum can be evaluated using Bluestein's technique \cite{Blu70}, which is related to the chirp z-transform. By noticing that $2 k n = n^2+k^2-(k-n)^2$ the Partial Fourier Transform above can be written as a convolution:
\begin{equation}
\widehat{F}(u_k) = \frac{D}{N} e^{i (\pi/\gamma) (k -N_B/2) }\sum_{n=0}^{N_A} e^{i \pi n N_B/(\gamma N_A)} f(x_n) e^{i \pi k^2/(\gamma N_A)} e^{-i \pi (n-k)^2/(\gamma N_A)}
\end{equation}
Consequently, the computation of the field in the image plane to an arbitrary sampling can be computed using a discrete convolution of sequences of length $N_A$. We direct the reader to the aforementioned references for details on these computation schemes. In order to use fast circular convolution algorithms, the incoming sequence needs to be zero-padded by factor of two, and the computational cost corresponds then to three FFTs of size $2 N_A$. 
When the field in the image plane is computed up to the Nyquist limit, the computational cost of the method is $20 N_A^2 \,\log_2 (N_A^2)$. When the size of the final array is $N_B \neq N_A$, partial convolution algorithms can be used, and they reduce the computational cost to $20 N_A^2 \,\log_2 (N_B^2)$ \cite{BS01}. 
In the case of large FT sizes ($N_B$ large) the fractional FT method has a lower cost than the MFT and the potential gain is of the order of $N_B/\log_2(N_B)$.

\section{Performance comparison and practical applications}\label{Sec:CompCosts}

\subsection{Computation costs for coronagraphic propagation}

We compare the computation costs of coronagraphic propagation from pupil to Lyot plane in the case of the FFT and semi-analytical methods.
As described before, coronagraphic propagation with FFTs requires pupil zero-padding by a coefficient $\gamma$. The array sizes are therefore $(\gamma N_A)^2$, where $N_A$ is the number of pixels across the pupil diameter and the number of flops involved in each FFT is $10 (\gamma N_A)^2 \,\log_2 (\gamma N_A)$.
The total number of operations for a propagation from pupil to Lyot plane is:
\begin{equation}
n(FFT)=20 (\gamma N_A)^2 \,\log_2 (\gamma N_A) + 6 (\gamma m)^2,\label{Eq:costTLPFFT}
\end{equation}
where we also include the multiplication by the focal plane mask ($6 (\gamma m)^2$ flops), which can be neglected in most cases of interest.

With the semi-analytical method, the size of the first FT is limited to the mask area, $N_B=\gamma m$ where $m$ is the mask size in resolution elements (we remind here that typically $m=5$ for a Lyot coronagraph and $m=1$ for a Roddier or DZPM).
The number of operations for the first MFT (Eq.~\ref{MFTcost}).
The second MFT transforms the $N_B\times N_B$ array into a $N_A\times N_A$ array and its cost is $n(MFT)=8 (N_A^2 N_B+ N_A N_B^2)-2N_A N_B-2 N_A^2$. 
 The propagation from pupil to Lyot plane includes the two MFTs, the multiplication by the focal plane mask ($6 N_B^2$ flops) and the subtraction of the result from the pupil ($2 N_A^2$ flops). The total cost of a semi-analytical propagation from pupil to Lyot plane is therefore:
\begin{equation}
n(SAM)=16 (N_A^2 \, \gamma \,m + N_A (\gamma\, m)^2) + 4 (\gamma\, m)^2 - 4N_A N_B,
\end{equation}
where the last two terms can be neglected in most cases of interest.
The relative computation cost between the classical FFT propagation and the semi-analytical method is therefore approximately:
\begin{equation}
\frac{n(FFT)}{n(SAM)}=\frac{5}{4} \frac{\gamma\,N_A \,\log_2(\gamma N_A)}{m N_A+ \gamma m^2},\label{Eq:costTLPMFT}
\end{equation}
neglecting the multiplication by the focal plane mask. This expression shows that there are always more operations with the FFT than with the semi-analytical method, and that the relative cost of the FFT increases with the padding factor $\gamma$ and the pupil size $N_A$.

\begin{figure}[htbp]
\center
\resizebox{0.75\hsize}{!}{\includegraphics{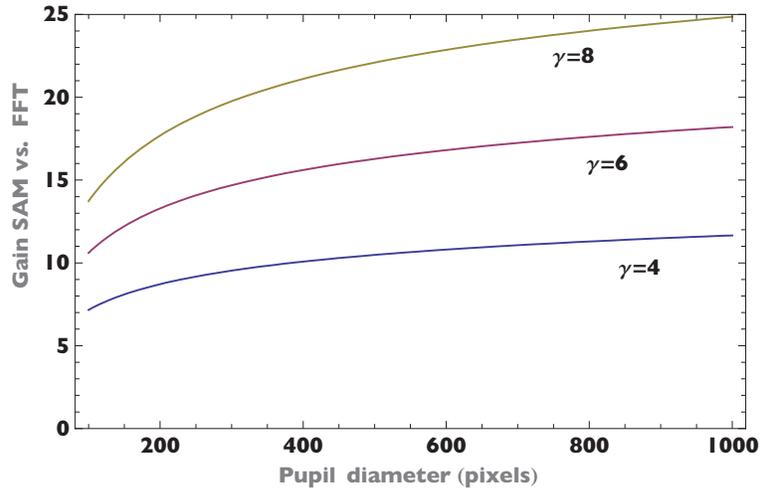}}
 \caption{Gain between SAM and FFT for a propagation from pupil to Lyot plane, as a function of the number of pixels across the pupil. The gain is based on the evaluation of the computation costs for both methods. We use common zero-padding coefficients $(\gamma=4,6,8)$ and a mask size $m=5\lambda/D$. This corresponds to the regime of simulations of Lyot coronagraphs for 8-meter class telescopes.}
 \label{fig:comparisonSAMFFT}
 \end{figure}

\begin{figure}[htbp]
\center
\resizebox{0.75\hsize}{!}{\includegraphics{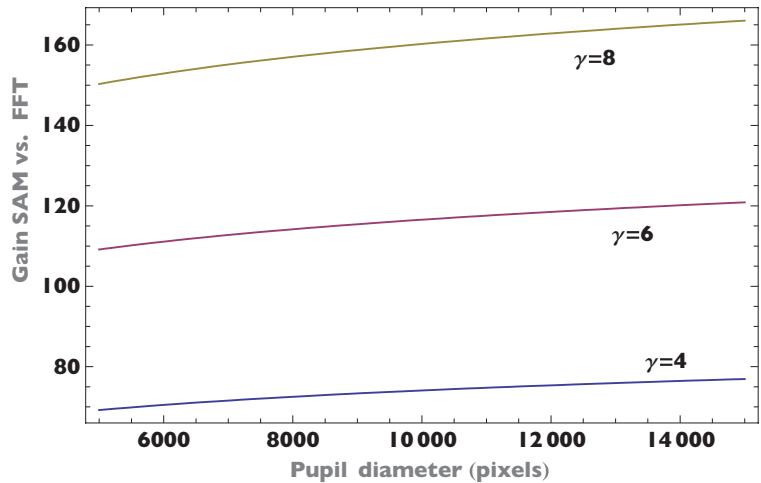}}
 \caption{Same as Fig.\ref{fig:comparisonSAMFFT}, but in the case of an ELT where a very large number of pixel is used in the pupil. }
 \label{fig:comparisonELT}
 \end{figure}

\begin{figure}[htbp]
\center
\resizebox{0.75\hsize}{!}{\includegraphics{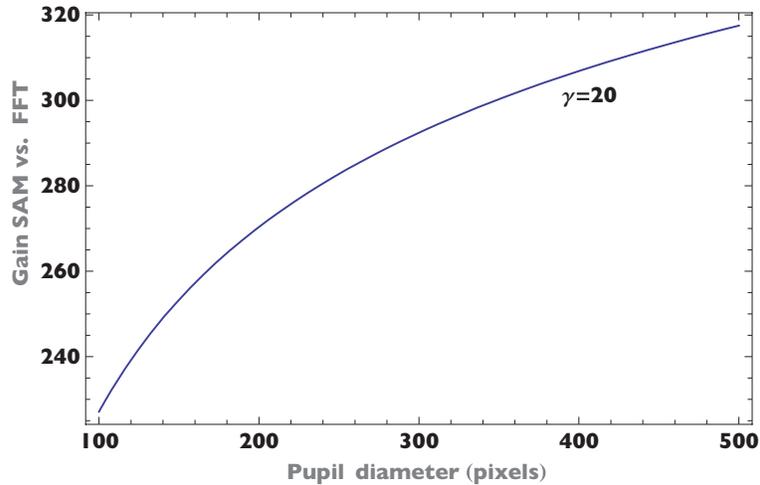}}
 \caption{Same as Fig.\ref{fig:comparisonSAMFFT}, but for a Roddier or Dual Zone coronagraph with a mask size of $m=\lambda/D$ and an oversizing $\gamma=20$. This corresponds to a maximum FFT size of $10000 \times 10000$ for the 500 pixel pupil.}
 \label{fig:comparisonRR}
 \end{figure}

In Fig.\ref{fig:comparisonSAMFFT}, Fig.\ref{fig:comparisonELT}, and Fig.\ref{fig:comparisonRR}, we illustrate the gain between the semi-analytical and FFT methods, using the ratio of their computation costs given in Eq.\ref{Eq:costTLPFFT} and Eq.\ref{Eq:costTLPMFT}.
We first give some results for a typical Lyot coronagraph simulation on an 8-meter class telescope, and two cases where the computation gain can be even larger: an ELT simulation with fine pupil sampling, or a phase mask coronagraph where very fine sampling is required in the focal plane.

\subsection{The case of direct imaging}

In this section we discuss briefly the case of direct imaging as a potential application of the MFT. 
Although in most cases the MFT requires more computations than a FFT, it is possible to identify a few cases where it can be more efficient.
This happens when the calculated area is limited and/or when a fine sampling is needed.
The MFT was first used by Give'on \textit{et al.} \cite{GKV05,GKV06} for the study of wavefront sensing and correction in high-contrast imaging with shaped pupils \cite{KVS03}.

\begin{figure}[htbp]
\center
\resizebox{0.75\hsize}{!}{\includegraphics{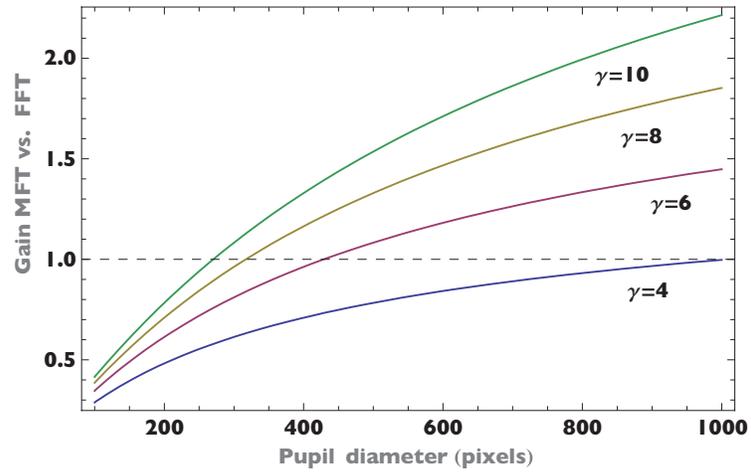}}
 \caption{Non-coronagraphic, direct imaging with FFT and MFT where the PSF is calculated to a field of view of 50 resolution element (25 in radius). Both MFT and FFT have similar costs with a slight advantage to MFTs when higher sampling is used.}
 \label{fig:comparisonDirectIm}
 \end{figure}

\begin{figure}[htbp]
\center
\resizebox{0.75\hsize}{!}{\includegraphics{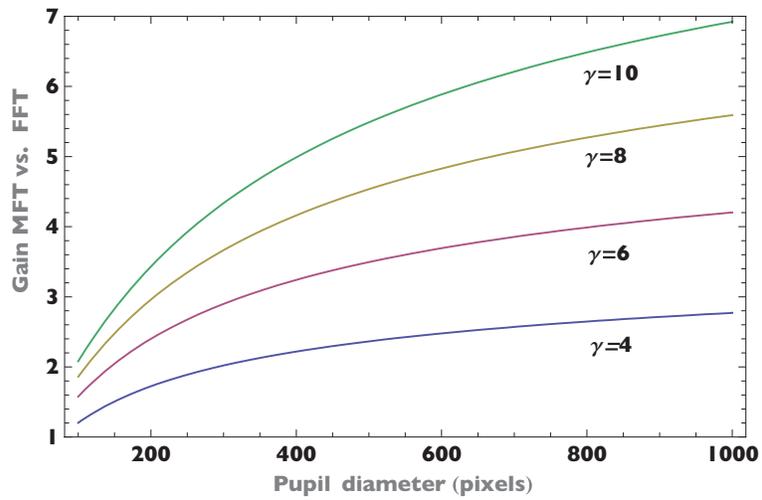}}
 \caption{Non-coronagraphic, direct imaging with FFT and MFT where the PSF is calculated to a field of view of 20 resolution element (10 in radius). For a limited field of view, the MFT can be significantly faster than the FFT. This can be interesting for coronagraphic and extreme adaptive optics simulations. }
 \label{fig:compDirectIm2}
 \end{figure}

We illustrate the case of a point spread function (PSF) calculation with a field of view ( FOV) of 50 resolution elements in Fig.\ref{fig:comparisonDirectIm}.
In this case, both techniques have similar costs with a slight advantage for one or the other depending on pupil size and sampling parameter $\gamma$.
If the PSF is only calculated to a FOV of 20 resolution elements, the MFT provides a gain by a factor of several over the FFT (Fig.\ref{fig:compDirectIm2}). In some applications where only the center of the PSF is needed, gains comparable to that of the coronagraphic cases can be obtained. 

\subsection{Practical applications and implementation}

\begin{figure}[htbp]
\center
\resizebox{0.75\hsize}{!}{\includegraphics{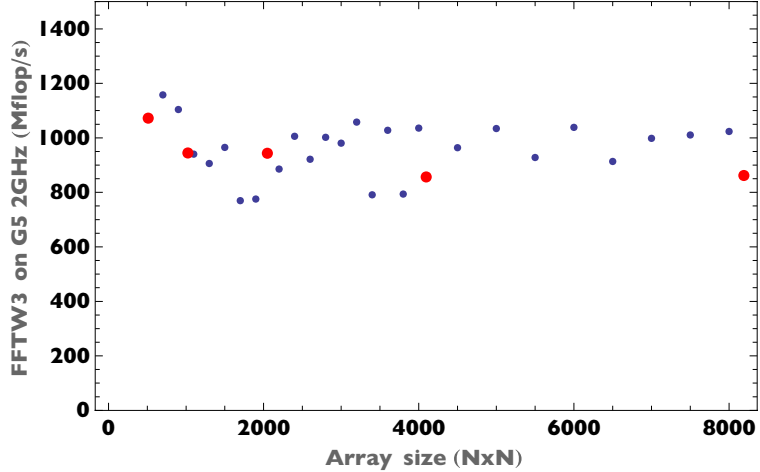}}
 \caption{Normalized speed expressed in Mflop/s using the FFTW3 package \cite{FFTW05} on a 2Ghz dual-processor 2003 Apple G5 with 4Gb RAM. The normalized speed is defined as the number of operations divided by the execution time: $10N^2\,\log_2(N)/T$ where $T$ is the time for one FFT (single-threaded, double precision complex arrays). The red dots correspond to power of twos arrays.This hardware delivers approximately 1Gflop/s over the range of array sizes, which translates for example into 2.3s for $4096\times4096$ transforms, and 10s for $8192\times8192$ transforms.}
 \label{fig:fftwflops}
 \end{figure}

In this section we discuss the actual performance of the semi-analytical method for a specific computer, a 2003 Apple G5 dual processor 2Ghz computer with 4Gb RAM. 
We measured the performance of this machine using the package FFTW (version 3) \cite{FFTW05} and report the results in Fig.\ref{fig:fftwflops}. We considered the case of double precision complex 2D arrays, which is appropriate for coronagraphic simulations. These results are consistent with those given on the FFTW website, but we extend them up to $8192\times8192$ FFT to verify that the speed does not depend strongly on the array size. Our calculations used the \verb"FFTW_PATIENT" option, which optimizes the FFT plan prior to the calculation of the FFT itself.
We used single-threaded FFTs although the FFTW packages provides the possibility of multi-threaded calculations for further speed improvement on multi-core machines. 

In order to evaluate the machine's performance for the MFT, we considered the Geekbench benchmark software \cite{geekbench} which reports 1.4 Gflop/s for single-threaded dot products and 2.6 Gflop/s for multi-threaded ones. 
It seems that dot products are slightly more efficient than the FFTW on this hardware (1.4 Gflop/s vs. 1 Gflop/s). This may be due to the easier implementation of the dot-product which takes direct advantage of the processor's vector engine.
An interesting aspect of the MFT is that it can take advantage of recent multi-core computers, with easy and efficient multi-threading of linear algebra calculations.

High level languages such at Mathematica or Matlab are well optimized for vector computations. For example, we obtain 2.5 Gflop/s with Mathematica for multi-threaded dot-products on double precision complex arrays of sizes between $1000\times1000$ and $10000\times10000$. This result is consistent with the benchmark results this machine (2.6 Gflop/s). 
In the case of a propagation from pupil to Lyot plane, the speed goes down to 1Gflop/s with Mathematica because the other operations involved in the semi-analytical method are not fully optimized with this language.

In Table 1, we compare the two methods for three pupil sizes ($256\times256$, $512\times512$, $1024\times1024$). 
$N_A$ and $N_B$ are the number of pixels used in the calculation, in the pupil and focal planes. We use the same number of pixels in the Lyot plane as in the pupil. We show the effect of the sampling improvement from $\lambda/4D$ to $\lambda/8D$, obtained by zero-padding the pupil with FFTs.
Note that for the semi-analytical method, the number of pixels $N_A$ corresponds to the pupil size itself, and that $N_B$ corresponds to the focal plane mask size. For the FFT, $N_A$ corresponds to the zero-padded pupil.
The timings given in the table correspond to the time for 2 FFTs calculated with FFTW3, neglecting the application of the focal plane mask and zero-padding time.
For the semi-analytical method, we give timings obtained with Mathematica as example. 

Coronagraphic simulations are usually performed with a $\lambda/6D$ to $\lambda/8D$ sampling, in order to keep the FFTs manageable. 
With the semi-analytical method, we can increase the precision significantly, and we typically use a $\lambda/10D$ or $\lambda/20D$ sampling for the focal plane mask, including gray pixel approximation. This corresponds to a $40\sim 50$ or $80\sim 100$ pixel diameter mask for a $4\sim 5$ resolution element mask. 

\begin{table}[htb]
\centering \caption{Performance comparison of FFT-based and semi-analytical methods (SAM). $N_A$ is the number of pixels across the pupil diameter, and $N_B$ is the size of the array in the focal plane. The FFT timings correspond to 2 FFTs calculated with FFTW3, and the SAM timings correspond to an actual propagation from pupil to Lyot plane, with Mathematica.}
\begin{center}
\begin{tabular}{ @{}| c | c | c | c | c | c | @{}}
\hline
Method &\multicolumn{2}{c |}{Pupil Plane} &\multicolumn{2}{c |}{Image Plane} & Time \\
\cline{2-5}
 &$N_A$&Array size &Sampling&Array size $(N_{B})$&\\
\hline \hline
FFT & $256$ & $1024$ &$\lambda/4 D$ & $1024$ & $0.22$ s \\ 

FFT & $256$ & $2048$ &$\lambda/8 D$ & $2048$ & $1.0$ s \\ 

SAM & $256$ & $256$ &$\lambda/8 D$ & $40$ & $0.06$ s \\ 
\hline \hline
FFT & $512$ & $2048$ &$\lambda/4 D$ & $2048$ & $1.0$ s \\ 

FFT & $512$ & $4096$ &$\lambda/8 D$ & $4096$ & $4.7$ s \\ 

SAM & $512$ & $512$ &$\lambda/8 D$ & $40$ & $0.19$ s \\ 
\hline \hline
FFT & $1024$ & $4096$ &$\lambda/4 D$ & $4096$ & $4.7$ s \\ 

FFT & $1024$ & $8192$ &$\lambda/8 D$ & $8192$ & $20.2$ s \\ 

SAM & $1024$ & $1024$ &$\lambda/8 D$ & $40$ & $0.70$ s \\ 
\hline
\end{tabular} \label{Tab:PerformancesComp}
\end{center}
\end{table}

The semi-analytical method offers a total flexibility in terms of sampling in both planes, and there is no particular limitation to the type of calculations. We present a few examples in Table 2, which would hardly be achieved with FFT propagations. For example, one can choose to increase the sampling in the pupil, with a modest $\lambda/8D$ sampling in the focal plane, in order to study the effect of segmentation of an ELT. We show an example of a $6000 \times 6000$ pupil array. This corresponds to calculations that are typically made on very large computer clusters using the classical method. This would correspond to $36000\times 36000$ FFTs with $\gamma=6$. 
It is also possible to oversample dramatically the occulting mask. This can be used to study the effect of the actual mask shape, such as ellipticity, edge roughness, fine alignment, or effects of atmospheric differential refraction. We give an extreme example with 1000 pixels in the pupil and a resolution of $\lambda/200D$. This calculation would corresponds to $200000 \times 200000$ FFTs. 

Another application is the simulation of Roddier or Dual Zone phase masks (DZPM), where the masks have a typical size of $\lambda/D$, making the sampling problem even more critical for FFTs. The optimization of DZPM \cite{SDA03}, which was obtained using the semi-analytical approach by integrating directly the Hankel transforms, can be studied by direct numerical simulations with our method. 

\begin{table}[htb]
\centering\caption{Performance of semi-analytical simulations for double precision complex arrays, with their possible applications. These calculations cannot be performed using the FFT method on commercial workstations, as they would correspond respectively to $200000\times 200000$, $25600\times 25600$, and $48000\times48000$ FFTs. }
\begin{center}
\begin{tabular}{| c | c | c | c | c | c | c |}
\hline
Method &\multicolumn{2}{c |}{Pupil Plane} &\multicolumn{2}{c |}{Image Plane} & Time & Application\\
\cline{2-5}
 &$N_{A}$ &Array size&Sampling&$N_{B}$& & \\
 \hline
SAM & $1000$ & $1000$ &$\lambda/200 D$ & $1000$ & $13.5$ s & Mask Tolerancing\\
\hline
SAM & $3200$ & $3200$ &$\lambda/8 D$ & $40$ & $6.1$ s & $4 $ pixel/spider on Gemini\\ 
\hline
SAM & $6000$ & $6000$ &$\lambda/8 D$ & $40$ & $22$ s & $5$ mm/pixel for 30 m ELT\\ 
\hline 
\end{tabular}
 \label{Tab:PerformancesDesk}
 \end{center}
\end{table}

\section{Conclusion}\label{sect:concl}

In this paper we introduced a semi-analytical method to calculate the numerical propagation through a Lyot-style coronagraph, without any additional approximation. We showed that FFTs are not appropriate for Lyot coronagraphy because of fundamental sampling limitations, and should be avoided for these calculations.
The semi-analytical method is derived straightforwardly from the analysis of the wave propagation, and requires the use of a Fourier transform method producing arbitrary sampling in a limited area. Several methods exist to perform such transforms, we suggest a matrix-based Fourier propagator that can be implemented efficiently in any language. 

For coronagraphic studies for ExAO on ELTs and eight meter class telescopes, the typical speed improvement based on computation costs is of at least a factor twenty to fifty compared to the classical FFT coronagraphic propagation method. In addition to speed, our semi-analytical method is particularly efficient in terms of memory, as it does not involve any zero padding, and typical memory reequirements are reduced by a factor of about 50 in our usage of the algorithm.
In the case of a full propagation from pupil to final focal plane, if the final FT is performed using a FFT, the semi-analytical method is approximately three times faster than a FFT-based propagation, since the propagation time is entirely dominated by the final FFT. 
If a reduced field of view is calculated in the final image, which is often the case in extreme adaptive optics coronagraphy, an MFT can be used in lieu of an FFT. For example a factor of 5 can be gained on the last FT for $N_A=400$ and $\gamma=10$, and the overall gain with the semi-analytical method is a factor of 15. 

With the semi-analytical method, the pupil and focal plane sampling are completely independent. This flexibility enables calculations that were hitherto impossible, or only possible on large and expensive computer clusters.
For example, the study of fine occulting mask features and alignment, or fine pupil structures like segmentation, is now achievable on a well-equipped but otherwise standard laptop, or on most commercial workstation sold today.

The MFT may also find interesting applications where a FT has to be calculated in a limited area. This was used by Give'on \textit{et al.} \cite{GKV05,GKV06} and new applications may be found in adaptive optics for anti-aliasing spatial filters \cite{PM04}, pyramid wavefront sensing \cite{Rag96}, etc. 

Our method also enables fast computation of optimal apodizers for APLCs, which involves an iterative algorithm and makes computing efficiency even more critical \cite{SPF07}. APLCs are one of the most promising concepts for high contrast imaging instruments on ELTs, and their study requires the optimization of the mask size according to various criteria, over broad spectral bandpasses. A complete exploration of the effects of chromaticity, segmentation and spiders on the coronagraph is now made possible with acceptable computing times. Our method could be utilized in several current efforts focussed on ground and spaced based coronagraphy.

\section*{Acknowledgements}
RS acknowledges partial support by a Michelson Postdoctoral Fellowship, under contract to the Jet Propulsion Laboratory (JPL) funded by NASA. JPL is managed for NASA by the California Institute of Technology and by a AMNH Kalbfleisch Fellowship.
This work has also been partially supported by the National Science Foundation Science and Technology Center for Adaptive Optics, managed by the University of California at Santa Cruz under cooperative agreement AST 98-76783. 
R. Vanderbei acknowledges support from the ONR (N00014-05-1-0206).
The authors thank Lisa Poyneer, Jim Fienup for interesting discussions, Russell Makidon and the reviewers for helpful comments.

\end{document}